\documentstyle[twocolumn]{jpsj}
\input epsf.sty
\title{Particle-Flux Separation and Quasi-excitations in Quantum Hall Systems}
\author{Ikuo {\sc Ichinose}\footnote{E-mail address:
ikuo@ks.kyy.nitech.ac.jp  } and
Tetsuo {\sc Matsui}$^{1,}$\footnote{E-mail address: 
matsui@phys.kindai.ac.jp} 
}
\inst{
Department of Electrical and Computer Engineering,
Nagoya Institute of Technology, Nagoya, 466-8555 Japan\\
$^1$ Department of Physics, Kinki University,
Higashi-Osaka 577-8502, Japan
}

\recdate{\today}

\abst{
The quasiexcitations of quantum Hall systems 
at the filling factor $\nu = p/(2pq \pm 1)$ are studied 
in terms of chargeon and fluxon introduced previously
as constituents of an electron at $\nu = 1/2$.
At temperatures $T < T_{\rm PFS}(\nu)$, 
the phenomenon so-called particle-flux separation takes place,
and chargeons and fluxons are deconfined to behave as quasiparticles.
Bose condensation of fluxons justify the (partial) cancellation of
external magnetic field.  Fluxons describe correlation holes, while
chargeons describe composite fermions. They contribute
to the resistivity $\rho_{xy} = h/(\nu e^2)$ additively.
}

\kword{quantum Hall effect, composite fermions,
gauge field theory, 
confinement-deconfinement transition}

\begin{document}
\sloppy
\maketitle


In the last decade, the quasi-excitations in a two-dimensional electron 
system
in a strong magnetic field have been intensively studied.
Currently, composite fermions\cite{Jain} (CF) for the filling factors
$\nu = p/(2pq + 1)\; (p,q;$ positive integers) and 
composite bosons\cite{ZHK,ZHK2,ZHK3} (CB)  for $\nu = 1/(2q +1)$  
play very important roles in a unified view of the quantum Hall 
effect (QHE), and the Chern-Simons (CS) gauge
theory is often used to describe them. 
The validity of these approaches 
relies upon the possibility that the CS fluxes attached
to each CF or CB cancel out the magnetic field, partly in the case of CF,
and completely in the case of CB (and CF at $\nu = 1/2$), 
as assumed in the mean-field theory (MFT).  

In ref.\cite{IM,IM2}, we considered the system of $\nu = 1/2$ and
argued that this cancellation really takes place at temperatures 
$T$ below a certain critical temperature $T_{\rm PFS}$ via the mechanism 
which we call particle-flux separation (PFS). 
In ref.\cite{IM2} we introduced the chargeon and fluxon 
operators to describe the charge and flux degrees of freedom of an electron,
respectively. They are confined within electrons by a U(1) gauge field
at $T > T_{\rm PFS}$, and  PFS is characterized as a deconfinement phenomenon
of the gauge dynamics of this gauge field.
The PFS is a counterpart of the charge-spin separation (CSS) \cite{CSS,CSS2} 
in high-$T_C$ cuprates, and justifies the CF picture of Jain\cite{Jain},
 Halperin et al.\cite{HLR}, and other works.

In recent years, some important studies of the CS gauge theory
of the QHE have appeared. 
Shankar and Murthy\cite{SM} studied the quasi-excitations 
in terms of a CF(CB) operator and the
longitudinal (or time)-component of the CS gauge field which becomes
dynamical because of the enlargement of the Hilbert space.
However, their CF(CB) operator {\em does not commute}
with the CS field and simple decoupling as they 
described {\em does not hold}.
Stern et al.\cite{SHOS} also
studied the transport properties of CFs at $\nu = 1/2$, but
the CS constraint is not respected.

Based on these studies, the treatments of CS constraint and the associated
ambiguity of independent variables certainly require further 
investigations in order to establish a consistent theory of CFs. 
In this letter we address this problem
by generalizing the chargeon and fluxon approach of ref.\cite{IM2} to
CFs at $\nu=p/(2pq \pm1)$. The lattice regularization and 
the second-quantized operators make it possible
to study PFS nonperturbatively
and  identify quasi-excitations in a consistent manner. Furthermore,
the transport properties are described quite neatly.

Let us consider the system of electrons on a two-dimensional lattice
in an external uniform magnetic field $B^{\rm ex}$ in the transverse 
direction, and start with the following CS representation of the electron
annihilation operator $C_{x}$ at the site $x$,\cite{IM,IM2}
\begin{eqnarray}
C_x=\exp\Big[2iq\sum_y \theta_{xy}\psi_y^\dagger\psi_y\Big]\psi_x,
\end{eqnarray}
where $\theta_{xy}$ is the multivalued angle function on a lattice,
and $\psi_x$ is the  fermionic annihilation operator of 
 a so-called CS fermion.  
The phase factor describes that each electron carries  
$2q$ ($q$: positive integer) units of CS flux quanta.
  The filling factor is given by $\nu=2\pi n/(eB^{\rm ex}a^2)$
$(\hbar = c = 1)$, where $n   \equiv \langle \psi_x^\dagger \psi_x\rangle $ 
is the average number of electrons or $\psi_x$'s {\it per site} 
(note that $C_x^{\dagger}C_x = \psi_x^\dagger\psi_x$),
and $a$ is the lattice spacing. 
We choose  $a \simeq \ell$ where 
$\ell = (e B^{\rm ex})^{-1/2}$ is the magnetic length. 
Hereafter, we set $a = 1$ in the formulae.
If $B^{\rm ex}$ is partly cancelled  by the CS fluxes 
uniformly as in MFT,  
each $\psi_x$ feels the residual constant magnetic field
$\Delta B \equiv  B^{\rm ex} - 4 \pi q n/e$. 
Then, FQHE takes place as the IQHE of $\psi_x$ in $\Delta B$ at
$1/\nu -2q =\pm1/p$ ($p$: positive integer), 
or $\nu = p/(2pq \pm1)$. Below we see that this idea is actually 
realized by PFS of chargeons and fluxons.

The Hamiltonian in terms of  $\psi_x$ is given by 
\begin{eqnarray}
H_\psi &=&-{1\over 2m}\sum_{x,j}\Big(\psi^\dagger_{x+j}
\exp[i(A_{xj}^{\rm CS}-eA_{xj}^{\rm ex}-ea_{xj})]
\psi_x
\nonumber\\
&&+{\rm H.c.}\Big) + H_{\rm int}(\psi_x^\dagger\psi_x),
\nonumber\\
A_{xj}^{\rm CS}&=& 2q\epsilon_{ji}
\sum_y \nabla_i\theta_{xy}\psi_y^\dagger\psi_y,
\label{Hpsi}
\end{eqnarray}
where $m$ is the effective electron mass, 
$j=1,2$ is the direction index (and the unit vectors), and
$A_{xj}^{\rm CS}$ is the CS gauge field,
its strength being
$B^{\rm CS}_x \equiv  \epsilon_{ij}\nabla_i A^{\rm CS}_{xj} 
 =4\pi q\psi^{\dagger}_x \psi_x$.
$H_{\rm int}$ represents replusive Coulombic interaction between 
electrons or $\psi_x$, 
and $A_{xj}^{\rm ex}$ is the electromagnetic (EM)  potential
for $B^{\rm ex},\ \epsilon_{ij}\nabla_iA_{xj}^{\rm ex} = B^{\rm ex}$.
To study the EM response functions, we  introduced an 
EM source potential $a_{xj}$.

Let us introduce the chargeon $\eta_x$ and fluxon $\phi_x$ operators
through
\begin{equation}
\psi_x=\phi_x\eta_x,
\label{CFoperator}
\end{equation}
which shows that $\psi_x$ is a composite of a chargeon and a fluxon.
We quantize the chargeon $\eta_x$ as a fermion, and 
the fluxon $\phi_x$ as a boson.\cite{hcb}
From eq.(\ref{CFoperator}), it is obvious that $\psi_x$ and $H_{\psi}$ 
are invariant
under the U(1) ``local gauge transformation", 
\begin{eqnarray}
(\eta_x,\phi_x)\rightarrow
(e^{i\alpha_x}\eta_x,e^{-i\alpha_x}\phi_x)\ {\rm for\ each\ } x.
\label{gaugesymmetry}
\end{eqnarray} 
To maintain equivalence  with eq.(\ref{Hpsi})
we impose the following local constraint:
\begin{equation}
\eta_x^\dagger\eta_x=\phi_x^\dagger\phi_x.
\label{const}
\end{equation}
Thus, the relations are $|0\rangle_{\psi}=|0\rangle_{\eta\phi},$ 
$\psi_x^{\dagger}|0\rangle_{\psi}=\eta_x^\dagger 
\phi_x^\dagger|0\rangle_{\eta\phi}$ at each $x$, where
$|0\rangle_{\eta\phi}=|0\rangle_{\eta} |0\rangle_{\psi}$ 
($\eta_x|0\rangle_{\eta}= \psi_x|0\rangle_{\psi} =0$). 

The electron operator is expressed  as
\begin{eqnarray}
C_x=\exp\Big[2iq \sum_y\theta_{xy} \phi^\dagger_y\phi_y\Big] \phi_x\eta_x.
\label{electron}
\end{eqnarray}
From eq.(\ref{electron}) an  electron 
is composed of a fluxon $\phi_x$,
a chargeon $\eta_x$, and  $2q$ units of CS flux quanta generated by
fluxons.
 This is illustrated in Fig.1.

$H_\psi$ of eq.(\ref{Hpsi}) is rewritten in terms of $\eta_x$ and $\phi_x$
as
\begin{eqnarray}
H_{\eta\phi}&=&-{1\over 2m}\sum_{x,j}
\Big(\eta_{x+j}^\dagger\phi_{x+j}^\dagger
W_{x+j}M_{x+j}M^\dagger_x W^\dagger_xe^{-iea_{xj}}\phi_x\eta_x\nonumber\\
&+&h.c.\Big)
-\sum_x (\mu_\eta\eta_x^\dagger\eta_x+\mu_\phi\phi_x^\dagger\phi_x)
+ H_{\rm int}(\eta_x^\dagger\phi_x^\dagger\phi_x\eta_x),
\label{Hetaphi}\nonumber\\
W_x&=& \exp\Big[2iq \sum_y\theta_{xy} (\phi^\dagger_y\phi_y -n)\Big],
\nonumber\\
M_x&=&\exp \Big[i\sum_y\theta_{xy}n\big(2q-\frac{1}{\nu}\big)\Big].
\label{W}
\end{eqnarray}	
We have added the terms with the chemical potentials $\mu_{\eta}, 
\mu_{\phi}$
to enforce $\langle \eta_x^{\dagger}
\eta_x \rangle = \langle \phi_x^{\dagger}\phi_x \rangle = n$.
(Note that $\psi_x^{\dagger}\psi_x = \phi_x^{\dagger}\phi_x
= \eta_x^{\dagger}\eta_x$.)

We are interested in the low-energy dynamics, particularly
how the local gauge symmetry (eq.(\ref{gaugesymmetry})) and the constraint 
(\ref{const}) are reflected there. 
We employ the path-integral  formalism and respect the constraint
(\ref{const}) by introducing the Lagrange multiplier field
$\lambda_x$. After decoupling $H_{\eta\phi}$
by introducing a complex auxiliary field $V_{xj}$ on the link $(x,x+j)$,
the Lagrangian is  expressed as
\begin{eqnarray}
L&=&-\sum_x \eta_x^\dagger(\partial_\tau-i\lambda_x-\mu_\eta)\eta_x
\nonumber\\
&-&\sum_x \phi_x^\dagger(\partial_\tau+i\lambda_x-\mu_\phi)\phi_x  
+{1 \over 2m}\sum_{x,j} \Big(V_{xj} J_{xj} + {\rm H.c.}\Big)
\nonumber\\
&-&{1\over 2m}\sum_{x,j} |V_{xj}|^2-H_{4}(\eta_x, \phi_x)
-H_{\rm int}(\eta_x^\dagger\eta_x
\phi_x^\dagger\phi_x),\nonumber\\
H_{4}
 &\equiv& \sum_{x,j }\Big({\gamma^2\over 2m}\phi_{x+j}^\dagger
\phi_x\phi_x^\dagger\phi_{x+j}
+{1\over 2m\gamma^2} \eta_{x+j}^\dagger\eta_x\eta_x^\dagger
\eta_{x+j}\Big)  \nonumber \\
J_{xj}&\equiv&\gamma \phi^\dagger_xW_{x}e^{ieca_{xj}}W^\dagger_{x+j}
\phi_{x+j}\nonumber\\ 
&&+\frac{1}{\gamma}\eta^\dagger_{x+j}M_{x+j}
e^{-ie(1-c)a_{xj}}M_x^\dagger\eta_x,
\label{L}
\end{eqnarray} 
where $\tau\in [0,\beta=1/(k_{\rm B} T)]$ is the imaginary time, 
and $\gamma$ is a 
parameter which measures the ratio of the masses of chargeon and fluxon.
 $c$ is an arbitrary constant,
 which appears in the EM charges of $\phi_x$ and $\eta_x$, 
\begin{eqnarray}
&&Q_{\phi} = c e,\ Q_{\eta} = (1-c)e. 
\label{EMcharge}
\end{eqnarray}
We shall discuss this important arbitrariness later.
From eq.(\ref{L}),  $A_{x0} \equiv \lambda_x$ and 
$A_{xj}$ of  $V_{xj} \equiv |V_{xj}| \exp(i A_{xj})$ can be regarded as
the time and spatial components of a U(1) gauge field $A_{x\mu}$.
The system has a full U(1) gauge invariance under
$A_{x\mu} \rightarrow A_{x\mu} + \nabla_{\mu} \alpha_x$ ($\nabla_0 \equiv 
\partial/\partial\tau$)
and eq.(\ref{gaugesymmetry}) with $\tau$-{\it dependent} $\alpha_x$.
There are no kinetic terms or Maxwell term of $A_{x\mu}$ in eq.(\ref{L}).
However, at low energies, $A_{x\mu}$ becomes dynamical as a result of 
``renormalization" (radiative corrections) by high-energy modes.
At low energies, there are two possible realizations of the gauge dynamics: 
(i) a deconfinement phase where the fluctuations of $A_{x\mu}$ are weak, and 
chargeons and fluxons are deconfined and behave as quasi-free particles, or 
(ii) a confinement phase where the fluctuations are
strong and chargeons and fluxons are confined into $\psi_x$, i.e., into
the original electrons.
The PFS is nothing but the deconfinement phenomenon (i),
as we shall see below.

To induce the PFS,
the repulsive Coulombic interaction $H_{\rm int}$ between 
electrons plays an important role. To clarify this, let us first
focus on its short-range (i.e., nearest-neighbor) 
part by setting 
$H_{\rm int}(\psi^\dagger_x\psi_x)
=g\sum\psi^\dagger_{x+j}\psi_{x+j}\psi^\dagger_x\psi_x$ with the
coupling constant $g\; (>0)$. It is natural to estimate $g$ as $g \simeq 
e^2/(\epsilon \ell)$, where $\epsilon$ is the dielectric constant.
Because $\psi^\dagger_{x+j}\psi_{x+j} \psi^\dagger_x\psi_x
=\eta^\dagger_{x+j}\eta_{x+j} \eta^\dagger_x\eta_x
=\phi^\dagger_{x+j}\phi_{x+j} \phi^\dagger_x\phi_x$ by 
eq.(\ref{const}), 
$H_{\rm int}$ may be rewritten at low energies as
\begin{eqnarray}
\hspace{-0.7cm} H_{\rm int} &=&
g_1\sum_{x,j} \eta^\dagger_{x+j}\eta_{x+j}\eta^\dagger_x\eta_x +
g_2\sum_{x,j} \phi^\dagger_{x+j}\phi_{x+j}\phi^\dagger_x\phi_x,
\label{g1g2}
\end{eqnarray}
where $ g_1 + g_2 = g$. Each term  
$H_{\rm int}$ or $H_4$ of eq.(\ref{L}) 
is difficult to respect nonperturbatively, but when
they are combined, one can treat them as irrelevant terms. 
In fact, we fix the values of $g_1, g_2, \gamma$ by requiring 
that $H_{\rm int}$ and $H_4$ cancel out,
$H_4 + H_{\rm int} = 0$, i.e.,
$g_1=1/(2m\gamma^2),g_2 =-\gamma^2/(2m)$.
This choice reflects
the idea that the fluxons and chargeons should behave
as freely as  possible since they are candidates for quasi-excitations
in the PFS state {\em at low energies}.

Let us put $V_{xj}=V_0 U_{xj}$ where $U_{xj}$  is a $U(1)$ variable
and  $V_0$ is the expectation value of $|V_{xj}|$ 
by ignoring its  fluctuations.
We discuss the estimation of $V_0$ later. 
 The  effective action $S_{\rm eff}$ of $A_{x\mu}$ at low energies 
is then obtained
by integrating out $\eta_x, \phi_x$.
We use the temporal gauge. At $T=0$, one can set $\lambda_x=0$.
However, at finite $T$, the zero modes of 
$\lambda_x(\tau),\ \theta_x \equiv \beta^{-1}\int d\tau \lambda_x$, 
remain as integration variables in general. 
Thus
\begin{eqnarray}
\int[d\eta][d\phi][d\theta] \exp\big(\int^\beta_0d\tau L \big) 
&=& \exp(- S_{\rm eff}).
\label{Seff}
\end{eqnarray}

To study PFS, we use the hopping expansion in powers of $V_0 U_{xj}$. 
The calculations are made straightforward by employing the 
single-site propagators
like $\langle \eta_x(\tau_1)\eta^\dagger_y(\tau_2)\rangle = \delta_{xy}
f_{\eta}(\tau_1 - \tau_2)$ as utilized in ref.\cite{IM,IM2}.
The $\theta_x$-integral in (\ref{Seff})  takes the form,
\begin{eqnarray}
&&\int [d\theta] \exp\Big(\sum_x \ln \frac{1+e^{\beta\mu_\eta +i \theta_x}}
{1-e^{\beta\mu_\phi -i \theta_x}} + O(V_0^2)
\Big).
\label{thetaint}
\end{eqnarray}
From this integrand, we find that the ground state of $\theta_x$
is given by $\theta_x=0$ (mod($2\pi$)).  
The term of $O(V_0^2) $ in the exponent of eq.(\ref{thetaint}) 
is expanded around $\theta_x=0$ as
 $-bV_0^2 \sum_{x,j}  (\nabla_j \theta_x)^2$ 
where $b $ is a positive function of $U_{xj}$. This assures that
$\theta_x=0$ is stable. The excitation modes of $\theta_x$ are massive
and the time component of $A_{x\mu}$ is screened, 
hence the perturbative calculations
which assume the small fluctuations of $\lambda_x$ are justified.
The constraint (\ref{const}) becomes irrelevant at low energies.
Therefore, we set $\lambda_x = 0$  in $L$  
to obtain 
\begin{eqnarray}
&&S_{\rm eff}= 
S_0+S_2 + O(V_0^4), \nonumber\\
&&S_2
=V_0^2\sum_{x,j}\Big[ \frac{\beta}{2m} -{n(1-n)\over 4m^2}
(\gamma^2 + \gamma^{-2})\beta^2
U^\dagger_{xj,0}U_{xj,0}\Big],\nonumber\\
&&U_{xj,0}\equiv \frac{1}{\beta} \int^\beta_0 d\tau 
U_{xj}(\tau).
\label{S2}
\end{eqnarray}

The properties of the quasi-excitations, i.e., whether 
PFS takes place or not,
depend on the behavior of $U_{xj}$. 
From $S_2$ of eq.(\ref{S2}), it is obvious that at large $\beta$, 
i.e., at low $T$,
 $U_{xj,0}$ dominates at $| U_{xj,0} | \simeq 1$ and 
the fluctuations of $A_{xj}$ are strongly suppressed.
In $O(V_0^4)$ of $S_{\rm eff}$, plaquette terms (magnetic
terms) like $U_{x2,0}U_{x+2,1,0}U^\dagger_{x+1,2,0}U^\dagger_{x1,0}$
appear, and their coefficients also become large at low $T$.
Therefore, at low $T$, $A_{xj}$ is in a deconfinement phase
and the PFS occurs.
Perturbative calculations with respect to $A_{xj}$ are justified.
The ``transition temperature" $T_{\rm PFS}$ is estimated 
by setting the coefficient of $| U_{xj,0} |^2$ in $S_2$ 
at unity\cite{IM,IM2},
\begin{eqnarray}
V_0^2(T_{\rm PFS})  \frac{n(1-n)}{4m^2 k_{\rm B}^2 T^{2}_{\rm PFS}}
(\gamma^2 + \gamma^{-2}) 
\simeq 1.
\label{TPFS}
\end{eqnarray}
The analysis developed in lattice gauge theory predicts that
the phase transition at $T_{\rm PFS}$ is smooth, as in CSS\cite{CSS,CSS2}, 
so our hopping expansion of $S_{\rm eff}$ in powers of $V_0 U_{xj}$ 
is justified
a posteriori. It corresponds to the Ginzburg-Landau
theory of global symmetry. 
This is in sharp contrast to most other studies of 
CS gauge theories working in the continuum.

%

Numerical estimation of $T_{\rm PFS}$ is given from eq.(\ref{TPFS}) 
for $\nu = 1/2$ 
by calculating $V_0^2(T)$ in a MFT of eq.(\ref{L}) obtained by setting
$\phi_x = \sqrt{n}, \lambda_x = 0$ \cite{TV} as
\begin{eqnarray}
T_{\rm PFS} &=& 4 \sim 4.5 {\rm K}\ \ \ {\rm for}  \ \ \ 
g = (0.1 \sim 1) \times \frac{e^2}{\epsilon \ell},
\label{TPFS2}
\end{eqnarray} 
where $ a = \ell$, $B^{\rm ex}=10[{\rm T}], m=0.067\, 
m_{\rm electron}, \epsilon=13$. Then  $\gamma=0.96 \sim 0.69$ and 
the masses of chargeon and fluxon at $T = 0$  are
$m_{\eta} \equiv \gamma  V_0^{-1} m = (6.5 \sim 4.7 ) m, 
m_{\phi} \equiv \gamma^{-1}  V_0^{-1} m  =(7.1 \sim 9.9 ) m.$
$T_{\rm PFS}$ of eq.(\ref{TPFS2}) seems consistent with the 
experimental results.\cite{PFSexp} 
The highest temperature $T_{\rm BC}$ at which 
FQHE  is observed is lower than $T_{\rm PFS}$ since FQHE is due to the 
Bose condensation of fluxons, as we shall see below.

We have obtained the above confinement-deconfinement phase transition (CDPT)
by using techniques of lattice gauge theory.
One may wonder if this transition survives in the ``continuum limit".
The CDPT at finite $T$  was first discovered
by Polyakov\cite{CD} and Susskind\cite{CD2} in  lattice gauge theory.
After that, more detailed investigations, including numerical studies
and renormalization-group (RG) analyses, 
confirmed the existence of this  CDPT in the continuum. 
The lattice models are regarded in these cases as effective models of 
RG, and the transition temperature is a RG-invariant quantity.

In the PFS states, one may 
neglect fluctuations
of $U_{xj}$ as the first approximation. 
Then, the ground state of electrons $|G\rangle_{C}$ is given by the product
$|G\rangle_{C}=|G\rangle_{\phi}|G\rangle_{\eta},$ where 
$|G\rangle_{\phi(\eta)} $ is the ground state of fluxons (chargeons). 
$|G\rangle_{\phi}$ describes the Bose condensate\cite{CB}. In the continuum
notation,
\begin{eqnarray}
&&\Psi_\phi(x_1,\cdots,x_N)\equiv
{}_\phi\langle 0 |\phi_{x_1}\cdots \phi_{x_N}|G\rangle_\phi  \nonumber  \\
&&=\prod_{i<j}|z_i-z_j|^{2q} \exp \Big[-\frac{1}{4\ell_\phi^2}
\sum_{j=1}^N |z_j|^2\Big],
\label{corH}
\end{eqnarray}
where $z_j$'s are the complex coordinates of $N$ fluxons, 
$\ell_\phi = (e B_{\phi})^{-1/2}\ (B_{\phi} \equiv \langle B^{\rm CS}_x 
\rangle /e = 4\pi q n /e)$. 
The CS factor $\exp [2iq \sum\theta_{xy}\phi^\dagger_y\phi_y]$ in 
eq.(\ref{electron}) produces a phase factor of $|z_i-z_j|^{2q}$,
changing $|z_i-z_j|^{2q}
\rightarrow (z_i-z_j)^{2q}$ in the electron wave function.
Thus we have
\begin{eqnarray}
\hspace{-0.8cm}&&\Psi_e(x_1,\cdots,x_N) \equiv
{}_C\langle 0 |C_{x_1}\cdots C_{x_N}|G\rangle_C  \nonumber  \\
\hspace{-0.8cm}
&=&\prod_{i<j}(z_i-z_j)^{2q}\ e^{-\sum |z_j|^2/(4\ell_\phi^2)}\cdot
{}_{\eta}\langle 0 |\eta_{x_1}\cdots \eta_{x_N}|G\rangle_\eta.
\label{elwave}
\end{eqnarray}

At $\nu=p/(2pq\pm 1)$, the {\it uniform} CS field generated by
the condensation of fluxons partly cancels uniform $B^{\rm ex}$. 
Chargeons feel the residual field $\Delta B = B^{\rm ex} - B_\phi 
= \pm 2\pi n/(ep)$,
and fill the $p$ Landau levels of $\Delta B$, giving rise to IQHE.
This observation obviously implies that the chargeons 
are nothing but Jain's CFs\cite{Jain}.
The wave function of $\eta$ in eq.(\ref{elwave}) 
is known for $p=1$ as  the Slater determinant,
\begin{eqnarray}
\hspace{-0.5cm}
{}_{\eta}\langle 0 |\eta_{x_1}\cdots \eta_{x_N}|G\rangle_\eta 
&=&\prod_{i<j}(z_i-z_j)\ e^{-\sum |z_j|^2/(4\ell_\eta^2)},
\label{etawave}
\end{eqnarray}
where $\ell_\eta = (e \Delta B)^{-1/2}$. Thus eq.(\ref{elwave})
becomes just the Laughlin's wave function for $\nu=1/(2q + 1)$.
(Note that $\ell^{-2} = \ell_\phi^{-2} + \ell_\eta^{-2}$.) 
For $p \neq 1$, one needs the wave function of IQHE.

At $\nu=1/(2q)\ (p = \infty)$, $\Delta B =0$, i.e., the uniform CS field 
generated by the fluxon condensate completely cancels out $B^{\rm ex}$,
thus chargeons behave as quasi-free fermions in zero
magnetic field.
Beyond the MFT, fluctuations of $A_{xj}$
mediating attractive interaction between chargeon and fluxons
may generate non-fermi-liquid behaviors.


Let us consider the EM transport properties of the PFS state.
The response functions of electrons are calculated from the 
effective action $S_{\rm EM}$ defined by 
\begin{eqnarray}
\int [dU]\exp(- S_{\rm eff}[a_{xj},U_{xj}]) = \exp(-S_{\rm EM}[a_{xj}]).
\label{SEM2}
\end{eqnarray}
In  the PFS states, fluctuations of the dynamical gauge field
$A_{xj}$ are small, so $S_{\rm eff}[a_{xj},U_{xj}]$
can be expanded in powers of $A_{xj}$ up to $O(A^2)$ as
\begin{eqnarray}
&&S_{\rm eff}[a_{xj},U_{xj}]=
\sum_{x,y,i,j}\Big[(A+cea)_{xi}\Pi^{ij}_{\phi; xy}(A+cea)_{yj}
\nonumber\\
&&+
(A+(1-c)ea)_{xi}\Pi^{ij}_{\eta; xy}(A+(1-c)ea)_{yj}\Big],
\label{Seff2}
\end{eqnarray}
where $\Pi^{ij}_{\phi(\eta)}$ is the polarization tensor
of $\phi_x (\eta_x)$.
$S_{\rm EM}[a_{xj}]$ is obtained by   integrating   over
 $A_{xj} (\in {\bf R})$  as 
\begin{eqnarray}
S_{\rm EM}[a_{xj}]&=& e^2 \sum a_{xi}\Pi_{xy}^{ij}a_{yj}, \nonumber\\
\Pi &=& (\Pi_{\phi}^{-1}+\Pi_{\eta}^{-1})^{-1},
\label{resfun}
\end{eqnarray}
where $\Pi$ is nothing but the response function of electrons. 
Then, we obtain the formula for the resistivity,
\begin{equation}
\rho = \rho_\eta+\rho_\phi,
\label{IL} 
\end{equation}
where $\rho = (e^{2} \Pi)^{-1},\ \rho_\eta$, and $\rho_\phi$ are the
$2 \times 2$ {\em resistivity tensor} of electrons,
chargeons, and fluxons, respectively.

The formula (\ref{IL})  does not depend on $c$ of eq.(\ref{EMcharge}).
In fact, $c$ 
expresses arbitrariness in choosing the reference state 
from which the relative EM charges (\ref{EMcharge}) are measured\cite{IMO}. 
A similar formula for $\rho $ is known for high-$T_c$ cuprates 
as the Ioffe-Larkin formula\cite{IL}.

What is the contribution to the electric transport from the fluxons?
In the CB theory for the FQHE\cite{ZHK,ZHK2,ZHK3}, each CB carries
$2q+1$-flux quanta and gives rise to $\rho_{xy} =(2q+1)h/e^2$.
The fluxons in the present formalism certainly contribute to  
$\rho_{xy}$ as do the CB, hence $\rho_{\phi\,xy}=2qh/e^2$.
Likewise,  $\rho_{\phi\,xx}=0$
because of the superfluidity of the fluxon condensate. 
On the other hand, the chargeons fill up the $p$ Landau levels 
of $\Delta B$ to 
contribute with $\rho_{\eta\,xy} = \pm h/(pe^2),\
\rho_{\eta\,xx} = 0$
as in the IQHE. Thus, from eq.(\ref{IL}), we obtain 
\begin{eqnarray}
\rho_{xy}\frac{e^2}{h} =2q \pm \frac{1}{p}=\frac{1}{\nu},
\ \ \rho_{xx} =0,
\end{eqnarray} 
which are actually observed in the experiments.
At $\nu=1/(2q)$, as a result of the condensation of 
fluxons, $\Delta B=0$ and the chargeon behaves as a Fermi liquid.
Therefore, $\rho_{xy}=\nu^{-1}(h/e^2)$ and $\rho_{xx} \neq 0$.
We shall discuss more details of the physics of
quasi-excitations.\cite{IMfull}

Finally, we comment on the role of Coulomb interaction.
Its short-range part enhances $T_{\rm PFS}$. 
In fact, eq.(\ref{TPFS}) shows that $T_{\rm PFS}$ increases
for larger $g$ (smaller $\gamma$) if $V_0(T)$ depends on $T$ weakly. 
On the other hand, the long-range part of Coulombic 
interaction may renormalize various effective 
parameters
such as the mass of chargeon and the strength
of repulsive interactions of fluxons, just 
as in a conventional Fermi-liquid theory.



\begin{figure}
  \begin{picture}(160,130)
    \put(0,0){\epsfxsize 160pt \epsfbox{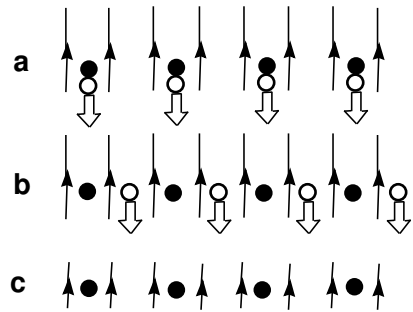}}
  \end{picture}
  \caption{
Illustration of PFS.
(a) Electrons in  magnetic field $B^{\rm ex}$. Thin arrows are $B^{\rm ex}$, 
black circles are chargeons $\eta_x$, white 
circles are fluxons $\phi_x$, and thick white arrows are CS fluxes. 
See eq.(\ref{electron}). (b) In PFS states, chargeons and 
fluxons dissociate. (c) In FQHE states,
fluxons form Bose condensate and the resulting uniform CS field 
cancels $B^{\rm ex}$ partly.
Chargeons feel the residual field $\Delta B$ (thin arrows).
}
\label{flux}
\end{figure}


\begin{thebibliography}{99}
\bibitem{hcb}
One may treat fluxons as  hardcore bosons\cite{IM,IM2} 
instead of canonical bosons. 
The results of PFS obtained below by the hopping expansion 
do not change qualitatively.
\bibitem{TV} 
$V_0$ starts to develop when the coefficient of $V_0^2$ in
 $S_2$ with $|U_{xj,0}| =1$  vanishes. The corresponding temperature is  
$T_V = 15 \sim 19{\rm K}$, which is larger than $T_{\rm PFS}$ of 
(\ref{TPFS2}), justifying the MFT for $V_0$.
The phase fluctuations of
$U_{xj}$ lower the transition temperature from $T_V$ down to $T_{\rm PFS}$.
\bibitem{CB} Strictly speaking, one needs certain amount of 
repulsive interaction among fluxons to obtain eq.(\ref{corH}).\cite{ZHK3}
This term may be supplied by relaxing the cancellation condition of the
short-range interactions.
The result (\ref{TPFS}) is almost unchanged.
\end{thebibliography}
\end{document}